*Research Article*

# Estimation Strategies for the Condition Monitoring of a Battery System in a Hybrid Electric Vehicle

S. A. Gadsden, M. Al-Shabi, and S. R. Habibi

*Department of Mechanical Engineering, McMaster University, Hamilton, ON, Canada L8S 4L7*

Correspondence should be addressed to S. A. Gadsden, gadsdesa@mcmaster.ca

This paper discusses the application of condition monitoring to a battery system used in a hybrid electric vehicle (HEV). Battery condition management systems (BCMSs) are employed to ensure the safe, efficient, and reliable operation of a battery, ultimately to guarantee the availability of electric power. This is critical for the case of the HEV to ensure greater overall energy efficiency and the availability of reliable electrical supply. This paper considers the use of state and parameter estimation techniques for the condition monitoring of batteries. A comparative study is presented in which the Kalman and the extended Kalman filters (KF/EKF), the particle filter (PF), the quadrature Kalman filter (QKF), and the smooth variable structure filter (SVSF) are used for battery condition monitoring. These comparisons are made based on estimation error, robustness, sensitivity to noise, and computational time.

## 1. Introduction

Condition monitoring is an essential process for fault detection and diagnosis. It involves monitoring system states or parameters over an operational period, where abnormal values or significant changes would indicate a fault. Quite often direct measurements of the states are not available due to limitations in design or cost. In these cases, state and parameter estimation techniques can be used for information extraction. Condition monitoring of systems allows proper maintenance to be scheduled, which helps reduce unscheduled downtime of manufacturing equipment, as well as the cost to repair damaged systems [1, 2]. An important area for condition monitoring is energy management for hybrid electric (HEVs) and battery electric vehicles (BEVs).

In general, HEVs have two power sources: a gasoline engine and an electric motor. In full hybrid vehicles, the engine and the motor can operate separately or simultaneously. The motor is used mainly during acceleration, startup, reverse mode, and in regenerative braking. A traction battery pack is used to provide power to the motor. It is recharged by a generator or during regenerative braking. The performance of an HEV is largely dependent on a balance between the gasoline engine and the electric motor, optimized with respect to fuel consumption based on vehicle conditions [3]. Many different types of control methods have been applied to balance the power and energy requirements of HEVs, including fuzzy logic [4–6], genetic algorithms [7], dynamic programming [8, 9], Pareto optimization [10], and intelligent mechanism designs [11]. These control strategies rely heavily on the availability of battery power to balance the operation of the gasoline engine versus the electric motor. The available battery power may be obtained from the state-of-charge (SOC) information [3, 12]. Further to the SOC, the battery state-of-health (SOH) is required in order to help determine whether a battery would fail subject to a certain load [13]. The SOC cannot be measured directly with electric signals, and as such it often, needs to be estimated through current and voltage relationships [3]. Poor estimation or control of the SOC may lead to improper charging conditions and can degrade the efficiency and reliability of the batteries [13]. Hence, proper condition monitoring of batteries plays a pivotal role in the optimization of HEV performance, as well as extending the lifetime and increasing the reliability of the batteries [14].

State and parameter estimation techniques are an integral part of condition monitoring and are used when direct measurements of the states are not available. One of the most commonly studied methods for estimation is the Kalman filter (KF) for linear systems and its extended form (EKF) for nonlinear systems [15–19]. Operating conditions such as battery SOC, power fade, capacity fade, resistance, and instantaneous available power have been estimated well using the EKF [17]. Other methods such as the sigma-point Kalman filtering (SPKF) and support vector machine (SVM) have also been used for condition monitoring and fault diagnosis [20–22].

In critical applications, such as automotive that require added safety and reliability, the choice of the estimation method is very important and should be selected based on the linearity of the system, performance, robustness or sensitivity to noise and computational difficulty and time. In relation to battery condition monitoring, this paper presents a quantitative and qualitative comparison of the following methods: the Kalman and extended Kalman filters (KF/EKF), the particle filter (PF), the quadrature Kalman filter (QKF), and the smooth variable structure filter (SVSF).

## 2. State and Parameter Estimation Techniques

State and parameter estimation is essential for sensing and information processing in model-based condition monitoring. Estimation theory involves information extraction by tracking changes in physical parameters or operational states of the system. This paper studies four popular strategies: the Kalman and the extended Kalman filters (KF/EKF), the particle filter (PF), the quadrature Kalman filter (QKF), and the smooth variable structure filter (SVSF).

*2.1. Kalman and Extended Kalman Filters.* Even after 50 years, the Kalman filter (KF) remains the most studied and one of the most popular tools used in state estimation [23–26]. It may be applied to linear dynamic systems in the presence of Gaussian white noise, and provides an elegant and statistically optimal solution by minimizing the mean-squared error. The impact that the KF has had on estimation and control problems is considered by some scientists and engineers to be one of the greatest achievement in engineering and signal processing [26]. It is a method that utilizes measurements linearly related to the states, and error covariance matrices, to generate a gain referred to as the Kalman gain. This gain is applied to the a priori state estimate, thus creating an a posteriori estimate. The estimation process continues in a predictor-corrector fashion while maintaining a statistically minimal state error covariance matrix for linear systems.

The following two equations describe the system dynamic model and the measurement model used in general for state estimation. Refer to Appendix A for a description of the nomenclature

$$\begin{aligned} x_{k+1} &= A_k x_k + B_k u_k + w_k, \\ z_{k+1} &= C_{k+1} x_{k+1} + v_{k+1}. \end{aligned} \quad (1)$$

The next five equations form the KF algorithm and are used in an iterative fashion. Equation (2) extrapolates the a priori state estimate, and (3) is the corresponding a priori error covariance. The Kalman gain may be calculated by (4), and is used to update the state estimate and error covariance, described by (5) and (6), respectively.

$$\hat{x}_{k+1|k} = A_k \hat{x}_{k|k} + B_k u_k, \quad (2)$$

$$P_{k+1|k} = A_k P_{k|k} A_k^T + Q_k, \quad (3)$$

$$K_k = P_{k+1|k} C_k^T \left[ C_k P_{k+1|k} C_k^T + R_k \right]^{-1}, \quad (4)$$

$$\hat{x}_{k+1|k+1} = \hat{x}_{k+1|k} + K_k [z_k - C_k x_{k+1|k}], \quad (5)$$

$$P_{k+1|k+1} = [I - K_k C_k] P_{k+1|k}. \quad (6)$$

The effects due to model uncertainties can have a large impact on the stability and performance of the KF [26, 27]. For nonlinear systems, the EKF may be used. It is conceptually similar to the KF process. The nonlinear system and measurement matrices are linearized according to their corresponding Jacobian, which is a first-order partial derivative. This linearization introduces uncertainties in the estimation process; such that overlooked nonlinearities in the system may cause the EKF to become unstable [26].

*2.2. Particle Filter.* The particle filter (PF) has many forms: Monte Carlo filters, interacting particle approximations [28], bootstrap filters [29], condensation algorithm [30], and survival of the fittest [31], to name a few. Compared to the KF, it is newer, being introduced in 1993. Since then, the PF has become a very popular method for solving nonlinear estimation problems, ranging from predicting chemical processes to target tracking. The PF takes the Bayesian approach to dynamic state estimation, in which one attempts to accurately represent the probability distribution function (PDF) of the values of interest [32]. The PDF contains all of the pertinent statistical information and may be considered as holding the solution to the estimation problem [32]. Essentially, the distribution holds a probability of values for the state being observed. The stronger or tighter the prediction PDF, the more accurate the state estimate.

The PF obtains its name from the use of weighted particles or "point masses" that are distributed throughout the PDF to form an approximation. These particles are used in an iterative process to obtain new particles and associated importance weights, with the goal of creating a more accurate approximation of the PDF. In general, as the number of implemented particles becomes large, the PDF becomes more accurate [32]. An important step in the PF is that of resampling, which eliminates particles with low weights and multiplies those with high weights [32]. This helps to avoid the degeneracy problem with the PF, which refers to only one particle having a significant importance weight after a large number of recursions. Furthermore, it also increases the accuracy of the PDF approximation by replicating particles with high weights. The sequential importance resampling (SIR) algorithm is a very popular form of the PF and may be summarized by (7) to (10). The first equation draws samples or particles from the proposal distribution.

$$x_k^{(n)} \sim \pi \left( x_k \mid x_{k-1}^{(n)}, y_k \right). \quad (7)$$

The next equation updates the importance weights up to a normalizing constant

$$\hat{\omega}_k^{(n)} = \omega_{k-1}^{(n)} \frac{p\left(y_k \mid x_k^{(n)}\right) p\left(x_k^{(n)} \mid x_{k-1}^{(n)}\right)}{\pi\left(x_k \mid x_{k-1}^{(n)}, y_k\right)}. \tag{8}$$

Next, the normalized weights are calculated for each particle

$$\omega_k^{(n)} = \frac{\hat{\omega}_k^{(n)}}{\sum_{i=1}^n \hat{\omega}_k^{(i)}}. \tag{9}$$

Finally, a constant known as the effective number of particles is calculated as shown in (10). Resampling is performed if the effective number of particles is lower than some design threshold

$$\hat{N}_{\text{eff}} = \frac{1}{\sum_{i=1}^n \left(\omega_k^{(i)}\right)^2}. \tag{10}$$

*2.3. Quadrature Kalman Filter.* Similar to the PF, the quadrature Kalman filter (QKF) is a type of Bayesian filter that is able to model dynamic processes which are nonlinear and subject to non-Gaussian noise. In 2007, it was proposed that a set of Gauss-Hermite quadrature points could be used to parameterize the PDF [33, 34]. When compared with the EKF, it was found that the QKF method provides a more accurate least-squares solution [33]. Figure 1 shows the PDF of a nonlinearly transformed Gaussian random variable. The mean of the EKF appears to be biased, and the covariance is obviously far from the true covariance. However, the QKF mean and covariance match the true values quite well, with the 5 point QKF working the best. The main drawback to this method is the fact that the number of terms used in the Gaussian sum grows exponentially, which means that more memory will be used over a longer period of time [33]. The process of this filter is similar to the PF, in the sense that it may be solved recursively using two stages (time and measurement update). Please refer to Appendix B for the full QKF algorithm.

*2.4. Smooth Variable Structure Filter.* In 2002, the variable structure filter (VSF) was introduced as a new predictor-corrector method used for state and parameter estimation [27, 35]. It is a type of sliding mode estimator, where gain switching is used to ensure that the estimates converge to true state values. An internal model of the system, either linear or nonlinear, is used to predict an a priori state estimate. A corrective term is then applied to calculate the a posteriori state estimate, and the estimation process is repeated iteratively. The SVSF was later derived from the VSF and uses a simpler and less complex gain calculation [36]. In its present form, the SVSF is stable and robust to modeling uncertainties and noise, given an upperbound on uncertainties [36]. The basic concept of the SVSF is shown in Figure 2. Assume that the solid line in Figure 2 is a trajectory of some state (amplitude versus time). An initial value is selected for the state estimate. The estimated state is pushed towards the true value until it reaches a subspace around the actual state trajectory, referred to as the existence subspace. Once the value enters the existence subspace, the estimated state is forced to remain within it and into switching along the system state trajectory [36].

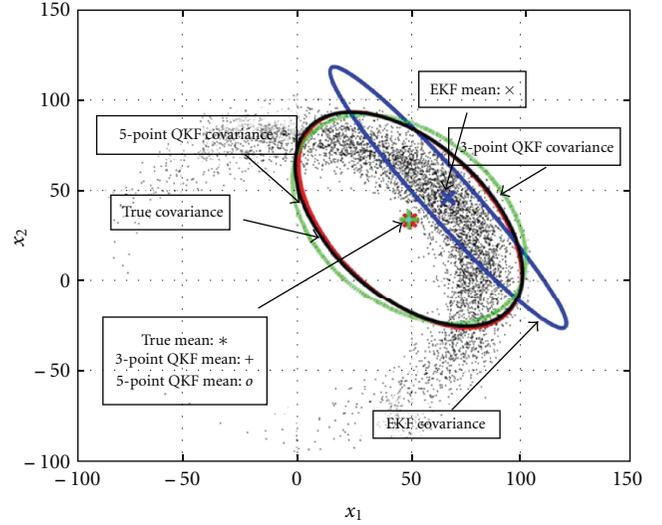

Figure 1: PDF of a Nonlinearly transformed gaussian random variable [33].

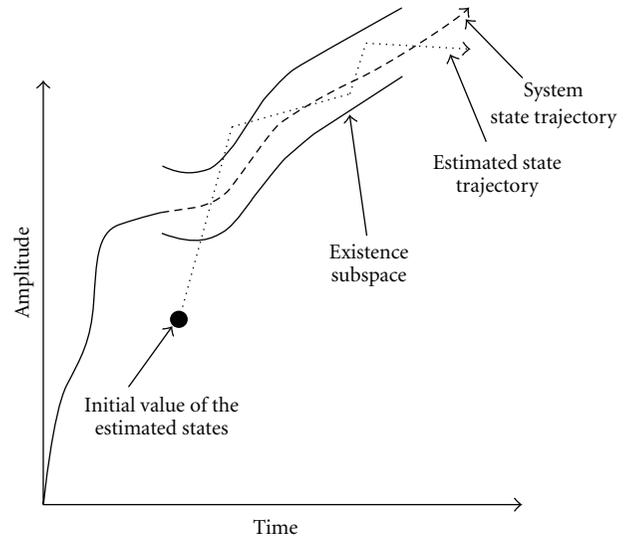

Figure 2: SVSF estimation concept [36].

The SVSF method is model based and applies to smooth nonlinear dynamic equations. The estimation process may be summarized by (11) to (16) and is repeated iteratively. An a priori state estimate is calculated using an estimated model of the system. This value is then used to calculate an a priori estimate of the measurement defined by (12). A corrective term, referred to as the SVSF gain, is calculated as a function of the error in the predicted output, as well as a gain matrix and the smoothing boundary layer width. The corrective term calculated in (13) is then used in (14) to find the a posteriori state estimate

$$\hat{x}_{k+1|k} = \hat{F}(\hat{x}_{k|k}, u_k), \quad (11)$$

$$\hat{z}_{k+1|k} = \hat{C}\hat{x}_{k+1|k}, \quad (12)$$

$$K_{k+1} = \hat{C}^{-1} \left| \left( \left| e_{z_{k+1|k}} \right|_{ABS} + \gamma \left| e_{z_{k|k}} \right|_{ABS} \right) \right|_{ABS} \circ \text{sat}\left( e_{z_{k+1|k}}, \Psi \right), \quad (13)$$

$$\hat{x}_{k+1|k+1} = \hat{x}_{k+1|k} + K_{k+1}, \quad (14)$$

$$e_{z_{k|k}} = z_k - \hat{z}_{k|k}, \quad (15)$$

$$e_{z_{k+1|k}} = z_{k+1} - \hat{z}_{k+1|k}. \quad (16)$$

Two critical variables in this process are the a priori and a posteriori output error estimates, defined by (15) and (16), respectively [36]. Note that (15) is the output error estimate from the previous time step and is used only in the gain calculation.

## 3. Condition Monitoring of a Battery System in a Hybrid Electric Vehicle

A variety of batteries have been studied in literature, most notably lead-acid and lithium-ion batteries [13, 15, 19, 20, 37]. Lead-acid batteries are the oldest type of rechargeable batteries, and are most commonly found in motor vehicles. Lithium-ion batteries are also a form of rechargeable battery, which contain lithium in its positive electrode (cathode). These batteries are usually found in portable consumer electronics (i.e., laptops or notebooks) due to particularly high energy-to-weight ratios, slow self-discharge, and a lack of memory effect (i.e., where a battery loses its maximum energy capacity over time) [16]. In recent years, lithium-ion batteries have slowly entered the hybrid electric vehicle market, due to the fact that they offer better energy density compared to standard batteries [38].

The operation of batteries may be studied by using the advanced vehicle simulator (ADVISOR), which was written in MATLAB and Simulink by the US Department of Energy and the National Renewable Energy Laboratory [39–41]. ADVISOR is used for the analysis of performance and fuel economy of three vehicle types: conventional, electric, and hybrid vehicles [39]. In 2001, the resistance-capacitance (RC) battery model was first implemented in ADVISOR [42]. The electrical model consists of three resistors ($R_e$, $R_c$, and $R_t$) and two capacitors ($C_b$ and $C_c$). The first capacitor ($C_b$) represents the capability of the battery to chemically store a charge, and the second capacitor ($C_c$) represents the surface effects of a cell [41]. The resistances and capacitances vary with changing SOC and temperature ($T$) [41]. ADVISOR offers two different datasets for the RC battery model: lithium-ion and nickel-metal hydride chemistries. For the purposes of this study, the lithium-ion chemistry was used in conjunction with the RC battery model. Figure 3 illustrates the equivalent circuit diagram of the RC model.

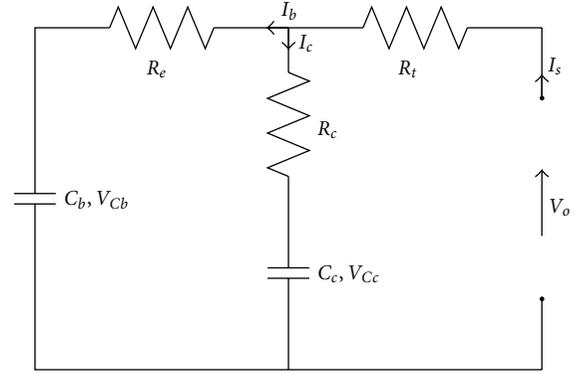

Figure 3: RC battery circuit diagram.

A standard model of a parallel hybrid electric vehicle referred to within ADVISOR as the Annex VII PHEV was used for this study. This model has been developed by the International Energy Agency (IEA), which is an international research community for the development and commercialization of hybrid and electric vehicles [43]. The model is based on data obtained from published sources and national (U.S.) research test data [39]. The battery system of the HEV represents the battery pack which stores energy on board the HEV. The system accepts a power request and returns the available power from the battery, as well as the SOC, voltage and current [39].

The nonlinear equations that describe the system may be derived from the RC battery model of Figure 3, as shown in Appendix C. For the purposes of condition monitoring of the battery, two voltages $V_{Cb}$ and $V_{Cc}$ as well as two capacitance parameters $C_b$ and $C_c$ need to be estimated. Should a fault exist in either of the battery capacitors, one would be able to determine this given the corresponding change in the parametric value. Further to the equations found in Appendix C, a discrete-time state space model of the Capacitor voltages may be defined as follows:

$$\begin{bmatrix} V_{Cb_{k+1}} \\ V_{Cc_{k+1}} \end{bmatrix} = \begin{bmatrix} -\dfrac{T_s}{C_b(R_e + R_c)} + 1 & \dfrac{T_s}{C_b(R_e + R_c)} \\ \dfrac{T_s}{C_c(R_e + R_c)} & -\dfrac{T_s}{C_c(R_e + R_c)} + 1 \end{bmatrix}_k \begin{bmatrix} V_{Cb_k} \\ V_{Cc_k} \end{bmatrix} + \begin{bmatrix} \dfrac{T_s R_c}{C_b(R_e + R_c)} \\ \dfrac{T_s R_e}{C_c(R_e + R_c)} \end{bmatrix}_k I_{S_k}. \quad (17)$$

For parameter estimation, (17) is used to formulate a state vector that would include parameters such that

$$x = \begin{bmatrix} V_{Cb} & V_{Cc} & C_b & C_c \end{bmatrix}^T. \quad (18)$$

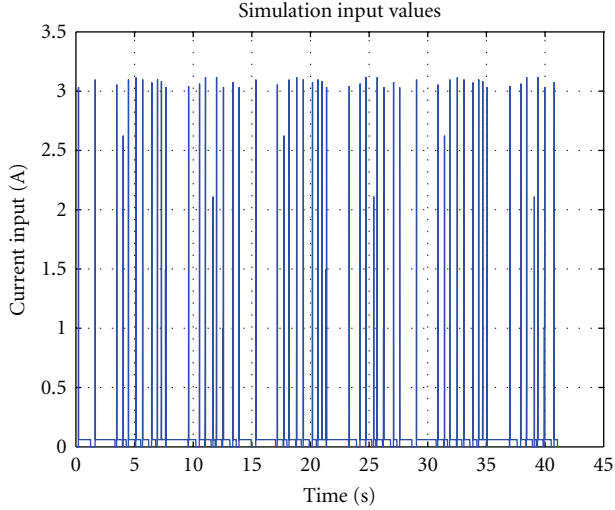

Figure 4: Supplied current input.

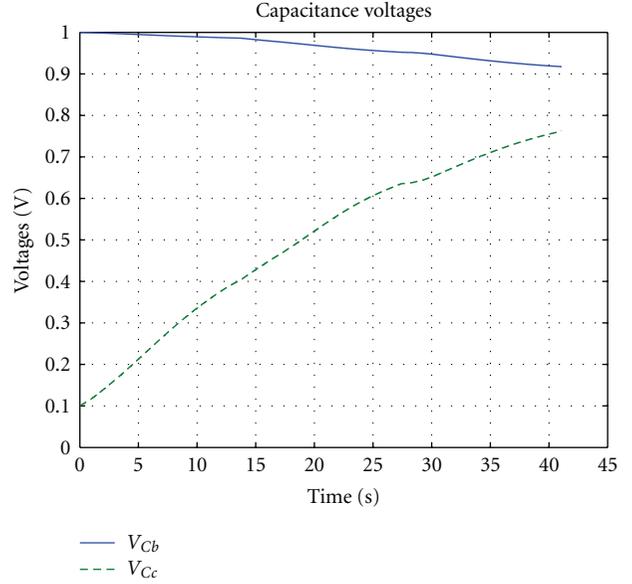

Figure 5: Capacitance voltages.

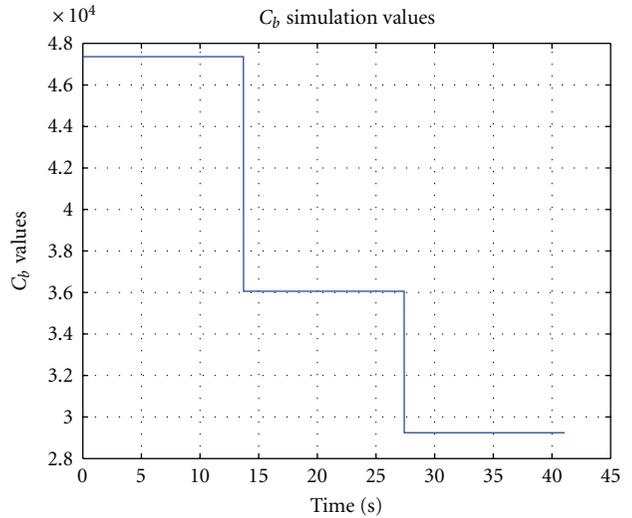

Figure 6: $C_b$ simulation values.

By rearranging (17), and further to Appendix C, the general model used for state and parameter estimation is obtained as

$$x_{k+1} = f(x_k, u_k) + w_k,$$
$$z_{k+1} = h(x_{k+1}, u_{k+1}) + v_{k+1}. \tag{19}$$

To implement the EKF method, the nonlinear system had to be linearized using a first-order Taylor series approximation. The linearized form of the system equation is provided in Appendix D.

## 4. Estimation Results

This paper presents a comparative performance for the application of the extended Kalman filter, the particle filter, the quadrature Kalman filter, and the smooth variable structure filter for condition monitoring of RC batteries. The study is conducted by simulation using the ADVISOR battery model. The model parameters are documented in [39].

Figures 4 and 5 illustrate the input current ($I_s$) and the output voltages $V_{Cb}$ and $V_{Cc}$ used in the simulation. The parameter values $C_b$ and $C_c$ are made to vary in order to simulate fault conditions as shown in Figures 6 and 7. Parameters $C_b$ and $C_c$ as well as the states $V_{Cb}$ and $V_{Cc}$ are estimated using the four methods described in Section 2.

A comparison and discussion of the results is provided using two simulated cases: one with noise and one with noise as well as modeling errors. Note that the sampling time used in the simulation was 0.01 seconds.

### 4.1. Extended Kalman Filter Results.
For linear dynamic systems in the presence of Gaussian white noise, the KF provides an elegant and statistically optimal solution by minimizing the mean-squared estimation error. The EKF is used for nonlinear problems. The following covariance matrices (error, process, and measurement, resp.) were used for the EKF and were obtained by trial-and-error:

$$P = \begin{bmatrix} 1 & 30 & 30 & 30 \\ 30 & 1 & 30 & 30 \\ 30 & 30 & 1 & 30 \\ 30 & 30 & 30 & 1 \end{bmatrix},$$
$$Q = R = \begin{bmatrix} 1 & 0 & 0 & 0 \\ 0 & 1 & 0 & 0 \\ 0 & 0 & 1 & 0 \\ 0 & 0 & 0 & 1 \end{bmatrix}. \tag{20}$$

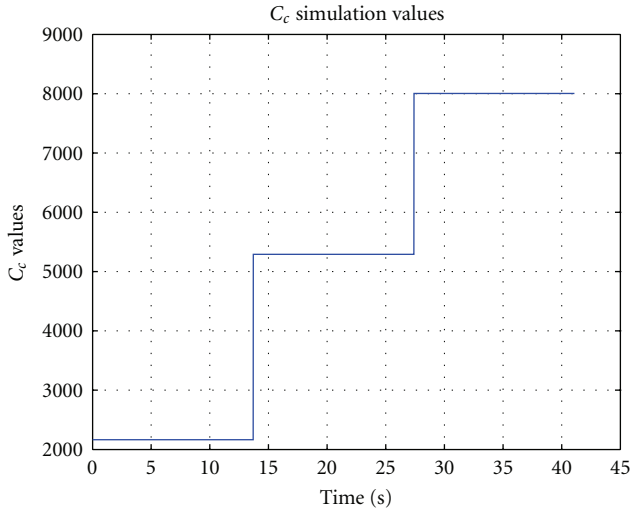

Figure 7: $C_c$ simulation values.

The EKF worked very well in the absence of modeling errors, as shown in Figure 8. Its performance degraded considerably when modeling errors were added, as illustrated in Figure 9.

As shown in Figure 10, the EKF was very sensitive to the selection of the initial conditions; however, it often recovered and performed very well. This is important to note because without selecting good starting points, the filter performed the worst. However, when good initial conditions were selected, the EKF yields the best results (based on RMSE and the assumption that there were no uncertainties in the model). When uncertainties (besides white measurement and process noise) were present in the filter model, in this example, the EKF became unstable (as shown in Figure 9) and failed to yield reasonable results. The simulation time for the EKF was one of the fastest, and compared to the other methods, it was computationally easy.

*4.2. Particle Filter Results.* For the PF method, a large number of particles (500) were used. Its application results are shown in Figures 11 and 12. The PF was able to estimate fairly well in the presence of noise. However, when modeling uncertainties were introduced, the PF had difficulty tracking the true voltages. As shown in Figure 12(b), after about 20 seconds, the PF was able to recover and provide a good estimate.

The PF provided satisfactory results. However, the technique had the highest RMSE (for the first case). That being said, when compared to the EKF, no Jacobian matrix had to be calculated to linearize the system matrix, as weighted particles were used instead. This may be an attractive feature if the system is too difficult to linearize. A large number of particles (which were required to accurately estimate the PDF), and the resampling feature of the PF, contributed to a slower computational time. When uncertainties were added to the filter model (as shown in Figure 12), the PF was able to recover slowly from modeling errors and after some time, accurately represent the first two states. There were significant errors in the third and fourth states; however this, as previously mentioned, may be attributed to measurements being only available for the first two states.

*4.3. Quadrature Kalman Filter Results.* For the QKF, the initial values of the covariance matrices were set to the identity matrix. The following results shown in Figures 13 and 14 were generated using the QKF based on the simulation setup. The QKF performed very well for both cases. However, like the previous methods, when uncertainties were introduced, it was unable to accurately track the two parameters associated with the battery ($C_b$ and $C_c$) as shown in Figure 14.

The QKF provided good results for the system. Like the PF, there was no need to calculate the Jacobian, since weighted quadrature points were used instead. One of the drawbacks of the QKF is the computational demand when a large number of states and quadrature points are used. This leads to a slower calculation time. For example, since four states were required, an array of 81 quadrature points was required ($3^4$). If a more accurate model was required, five quadrature points per state could be used. However, this would further increase the computational time. For this system, it was found that increasing the number of quadrature points had a negligible effect on the accuracy. Note that when uncertainties were added to the filter model, the QKF worked extremely well. However, the slow computation time is its main hindrance.

*4.4. Smooth Variable Structure Filter Results.* The constant diagonal matrix ($\gamma$) was set to 0.4, and the smoothing boundary layer thicknesses ($\psi$) were, respectively, set to $1 \times 10^{-3}$ and $1 \times 10^{-2}$ for the two states and the two parameters, respectively. The results are shown in Figures 15 and 16, and were generated using the SVSF based on the simulation setup. The SVSF method is very robust to noise and modeling uncertainties, as demonstrated by the accurate estimation in both cases.

The SVSF yielded very good results for condition monitoring of the battery system. Similar to the previous two methods, it did not require calculation of the Jacobian matrix, and also required a time delay (signal extraction for the last two states). The SVSF simulation time was just as fast as the EKF, and the results were similar for the case with noise. When errors were added to the filter model, the SVSF worked the best (in terms of RMSE). Although there were still large estimation errors for the third and fourth states, they were finite and stable. It is important to note that the SVSF was robust and not sensitive to changes in the filter parameters and the initial conditions.

*4.5. Summary of Results.* The results of the simulations were compared based on estimation error, robustness, sensitivity to noise, and computational difficulty and time. As shown in Table 1, the EKF performed best in terms of estimation error, in the case of only noise (referred to as Case (1)). In the presence of modeling uncertainties (Case (2)), the EKF's performance severely degraded. The SVSF yielded the most accurate estimation for the case involving modeling uncertainties. Both the PF and QKF performed moderately

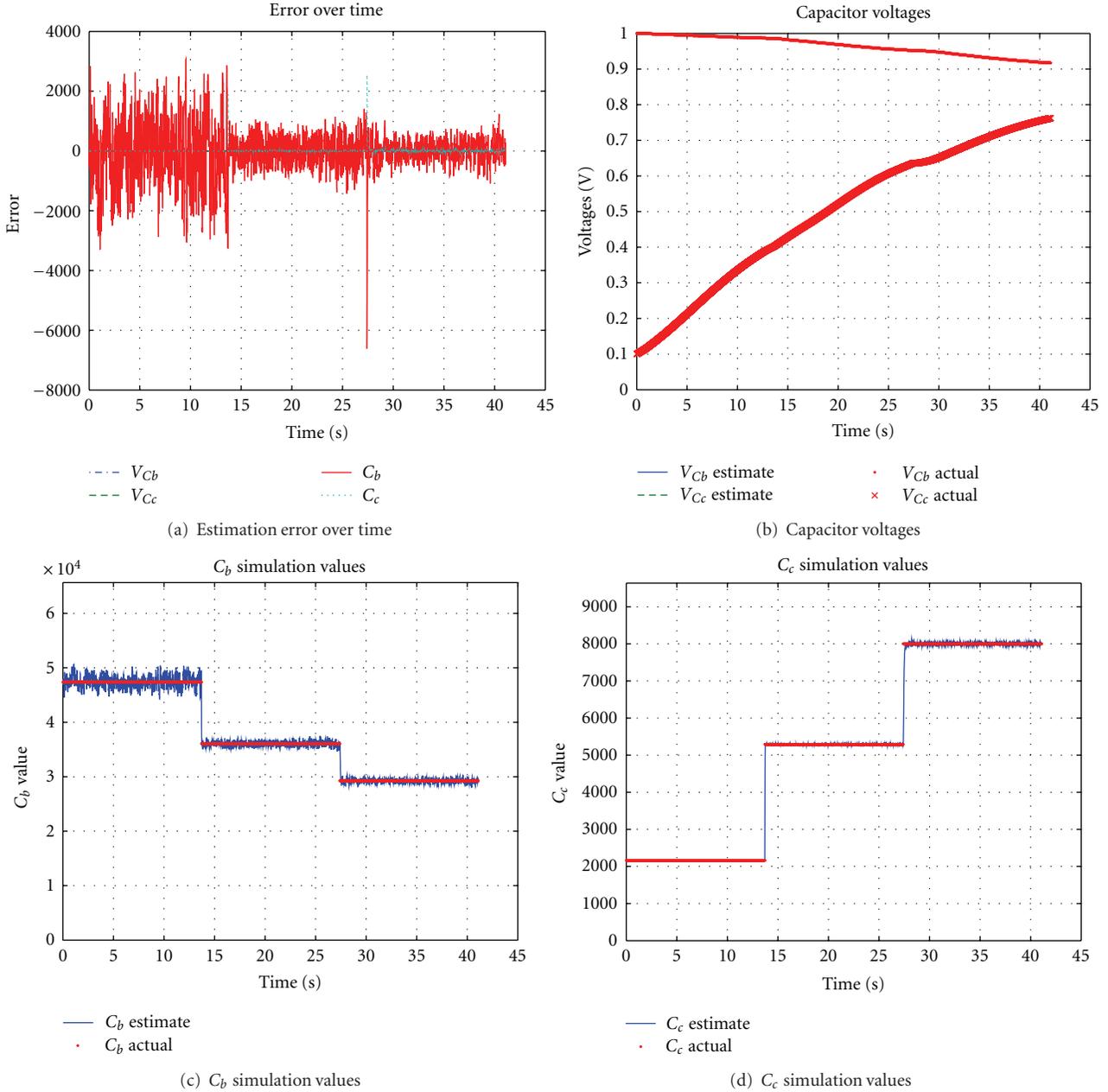

Figure 8: Extended Kalman filter simulation results (only noise).

well, but at a cost of computational time, as shown in the last row of Table 1.

The error in Table 1 refers to the root mean-squared error (RMSE), calculated as follows:

$$\text{RMSE} = \frac{1}{N}\sqrt{\sum_{i=1}^{N}(x_i - \hat{x}_i)^2}. \qquad (21)$$

Table 2 gives a quantitative comparison of the four filters. The robustness was based on observations made on varying the noise levels and considering its impact on model errors. The sensitivities were determined by tuning the filters various parameters. Computation (simulation) time was determined for each method and may be compared on a relative basis.

## 5. Conclusions

This paper discussed the application of condition monitoring to an RC battery system, typically found within a hybrid electric vehicle. State and parameter estimation techniques are important as they are responsible for providing accurate estimates of the states when reliable measurements are unavailable, and hence ensure successful condition monitoring. A comparative study was presented in which the Kalman and the extended Kalman filters (KF/EKF), the particle filter (PF), the quadrature Kalman filter (QKF), and the smooth variable structure filter (SVSF) were applied for condition monitoring. These estimation methods were compared based on estimation error, robustness, sensitivity to noise, and computation time. Table 3 summarizes the results of the comparison. For the case without modeling errors, the EKF worked the best in terms of RMSE and computational speed. When modeling errors were present, which is common in physical applications, the SVSF was shown to work significantly better than the other methods in terms of stability and RMSE.

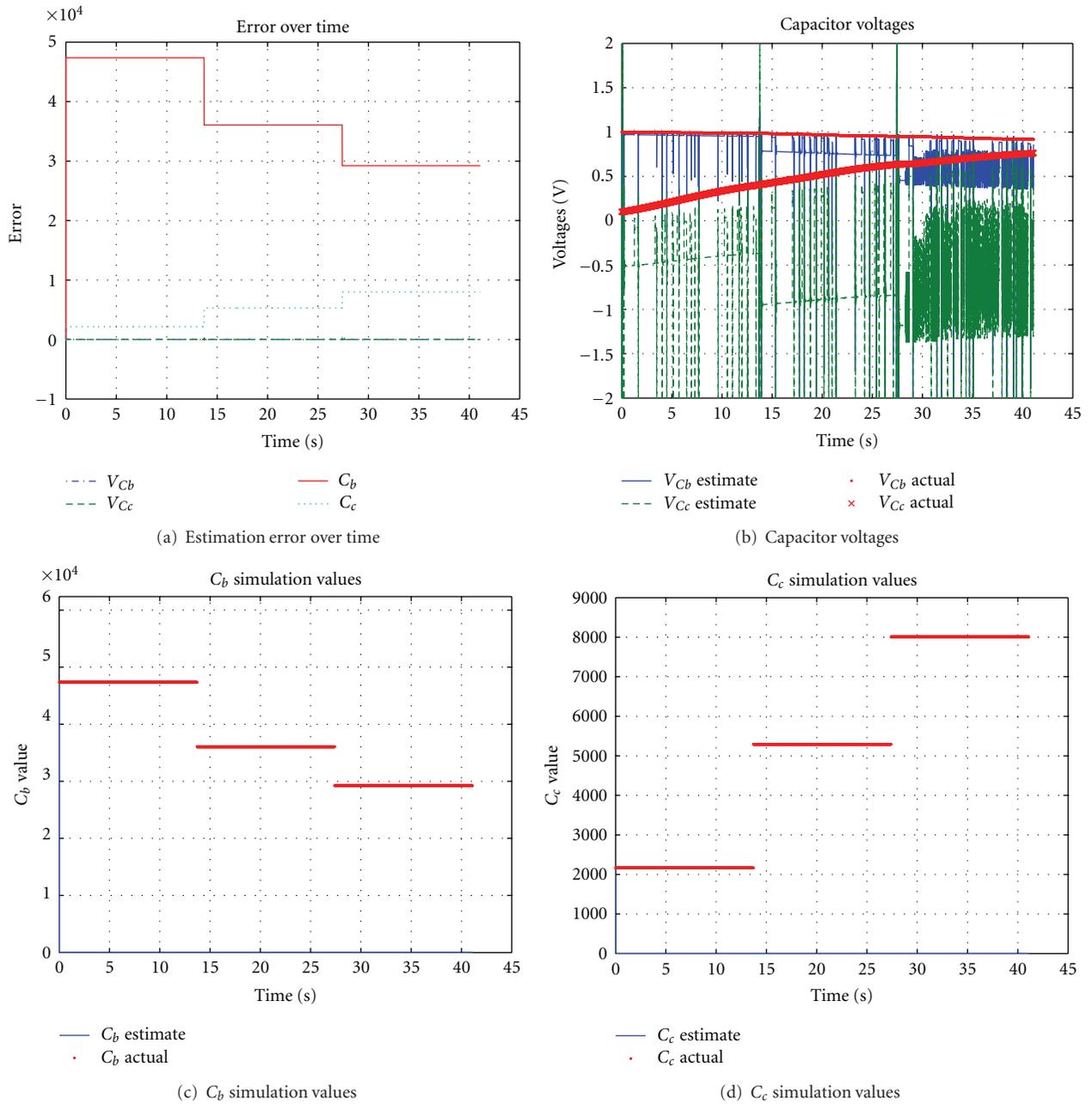

Figure 9: Extended Kalman filter simulation results (including errors).

Table 1: Quantitative summary of results.

| States and parameters | | EKF | | PF | | QKF | | SVSF | |
|---|---|---|---|---|---|---|---|---|---|
| | | Case (1) | Case (2) | Case (1) | Case (2) | Case (1) | Case (2) | Case (1) | Case (2) |
| RMSE | $V_{Cb}$ | 2.25E − 09 | 5.61E − 01 | 2.34E − 04 | 9.50E − 03 | 2.07E − 05 | 4.19E − 05 | 2.82E − 07 | 1.24E − 08 |
| | $V_{Cc}$ | 1.10E − 08 | 9.23E − 02 | 2.95E − 04 | 1.30E − 03 | 2.07E − 05 | 4.19E − 05 | 6.58E − 07 | 9.62E − 09 |
| | $C_b$ | 12.02 | 597 | 41.19 | 1196 | 28.07 | 1196 | 18.48 | 1195 |
| | $C_c$ | 1.40 | 89 | 2.19 | 177 | 1.20 | 177 | 0.73 | 177 |
| Simulation time (sec) | | 0.38 | 0.44 | 3.27 | 3.34 | 8.35 | 8.42 | 0.39 | 0.39 |

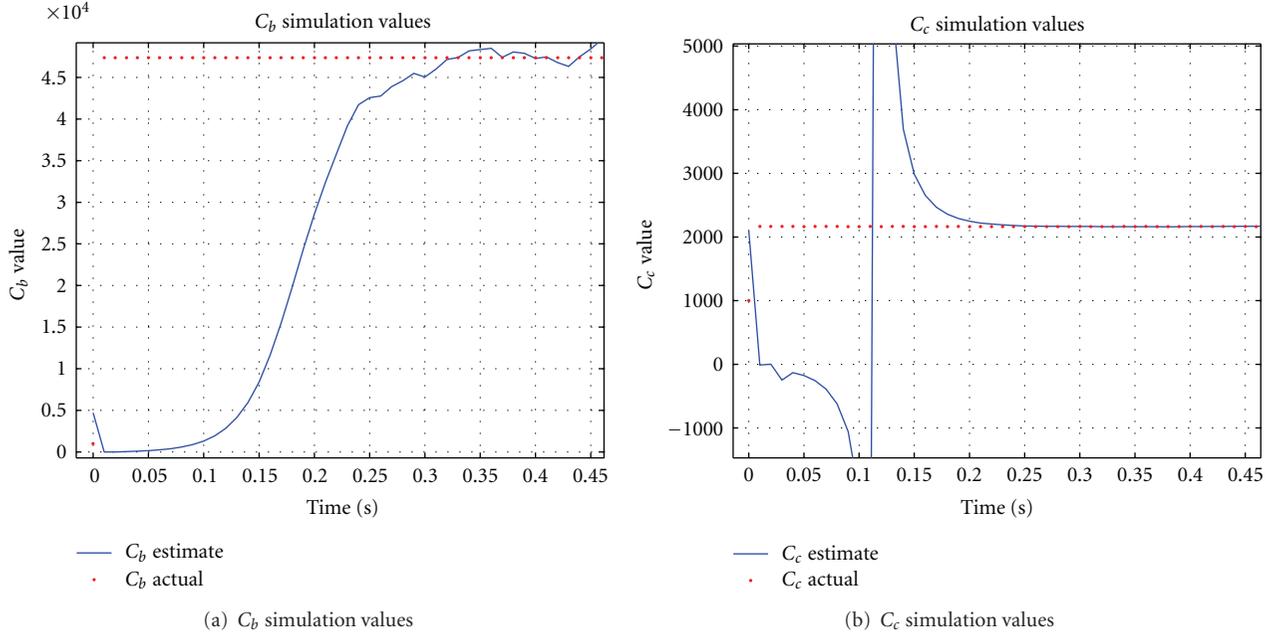

Figure 10: EKF Simulation results with poor initial conditions (only noise).

Table 2: Qualitative comparison of the filters.

| Characteristic | EKF | PF | QKF | SVSF |
|---|---|---|---|---|
| Robustness | Little | Medium | Good | Good |
| Sensitivity to Uncertainties | High | High | Low | Low |
| Sensitivity to Filter Parameters | High | Low | Low | Low |
| Computation Time | Fast | Slow | Very Slow | Fast |

Table 3: Ranking of the state estimation techniques.

| Ranking | Case (1): Only noise | Case (2): Noise and model errors |
|---|---|---|
| 1 | EKF | SVSF |
| 2 | SVSF | QKF |
| 3 | QKF | PF |
| 4 | PF | EKF |

# Appendices

## A. List of Nomenclature

$A$: System matrix
$B$: Input matrix
$C$: Output matrix
$e$: State estimation error
$f$: Nonlinear system or process equation
$h$: Nonlinear measurement or output equation
$k$: Time step index
$K$: Gain value (KF, QKF, or SVSF)
$m$: Number of measurements
$n$: Number of states
$N_{eff}$: Effective number of particles
$P$: Error covariance matrix
$Q$: System noise covariance matrix
$R$: Measurement noise covariance matrix
Sat: Saturation function
$t$: Simulation time
$T_s$: Sampling time
$u$: Input
$v$: Measurement noise
$w$: System noise
$w_l$: Quadrature point weight

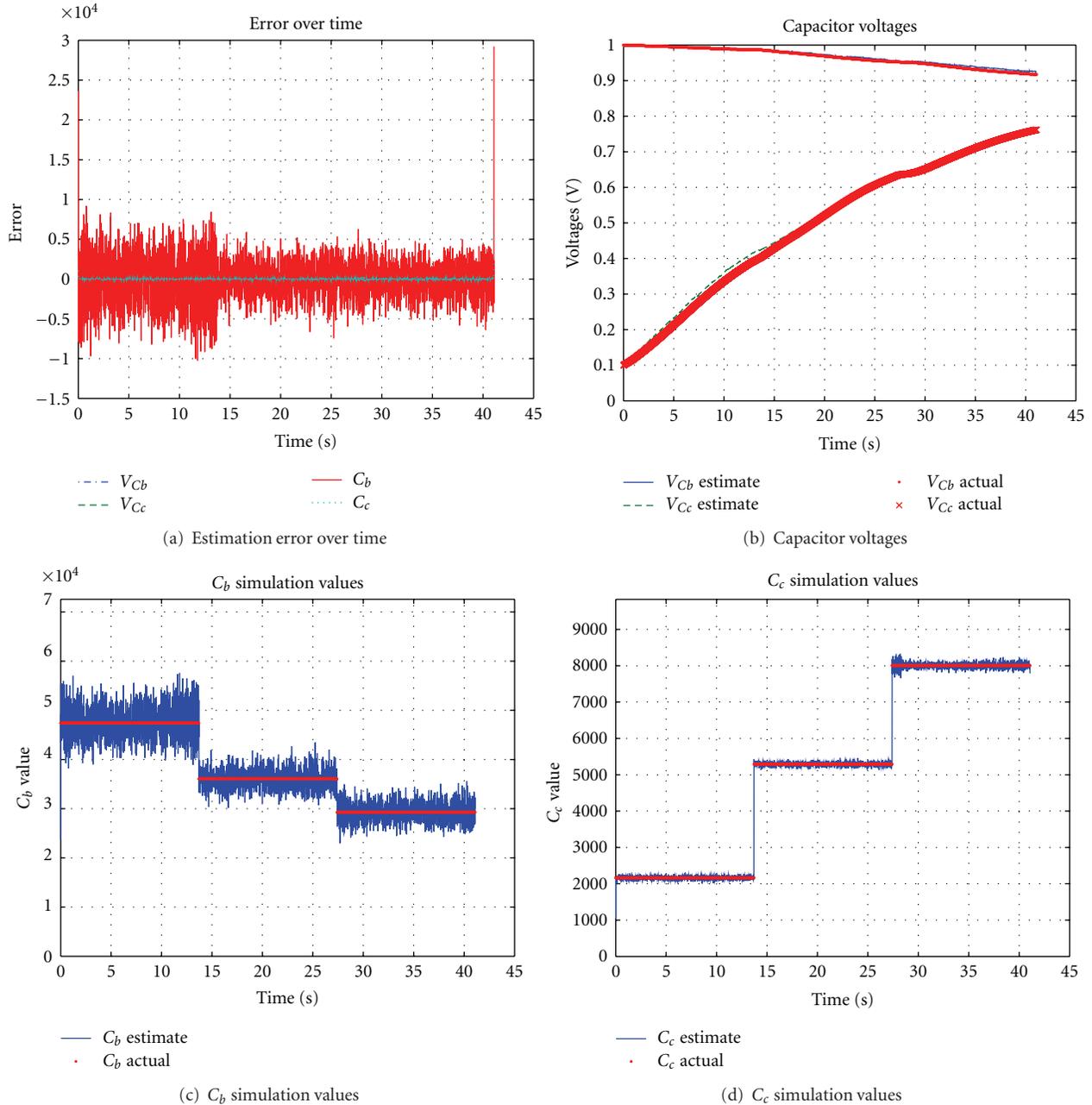

Figure 11: Particle filter simulation results (only noise).

- $x$: System states
- $y$: Observed measurement or output
- $z$: Measurement output
- $\gamma$: Constant diagonal gain matrix with elements having values between 0 and 1
- $\omega$: Particle weight
- $\Psi$: Smoothing boundary layer
- $\pi$: Probability distribution
- $\tau$: Sampling time
- $\hat{}$: Denotes an estimated value
- $\sim$: Denotes an error value
- $\cdot$: On top of a parameter denotes a time derivative.

Furthermore, note that subscript $k+1 \mid k$ refers to an a priori time step and that the subscript $k+1 \mid k+1$ refers an a posteriori time step. A superscript of $T$ denotes a matrix transpose.

## B. Quadrature Kalman Filter Algorithm

The following is a summary of the QKF algorithm, directly as presented in [33, 34]. This process was used to obtain the presented results. There are two updates at each step: the time update and the measurement update.

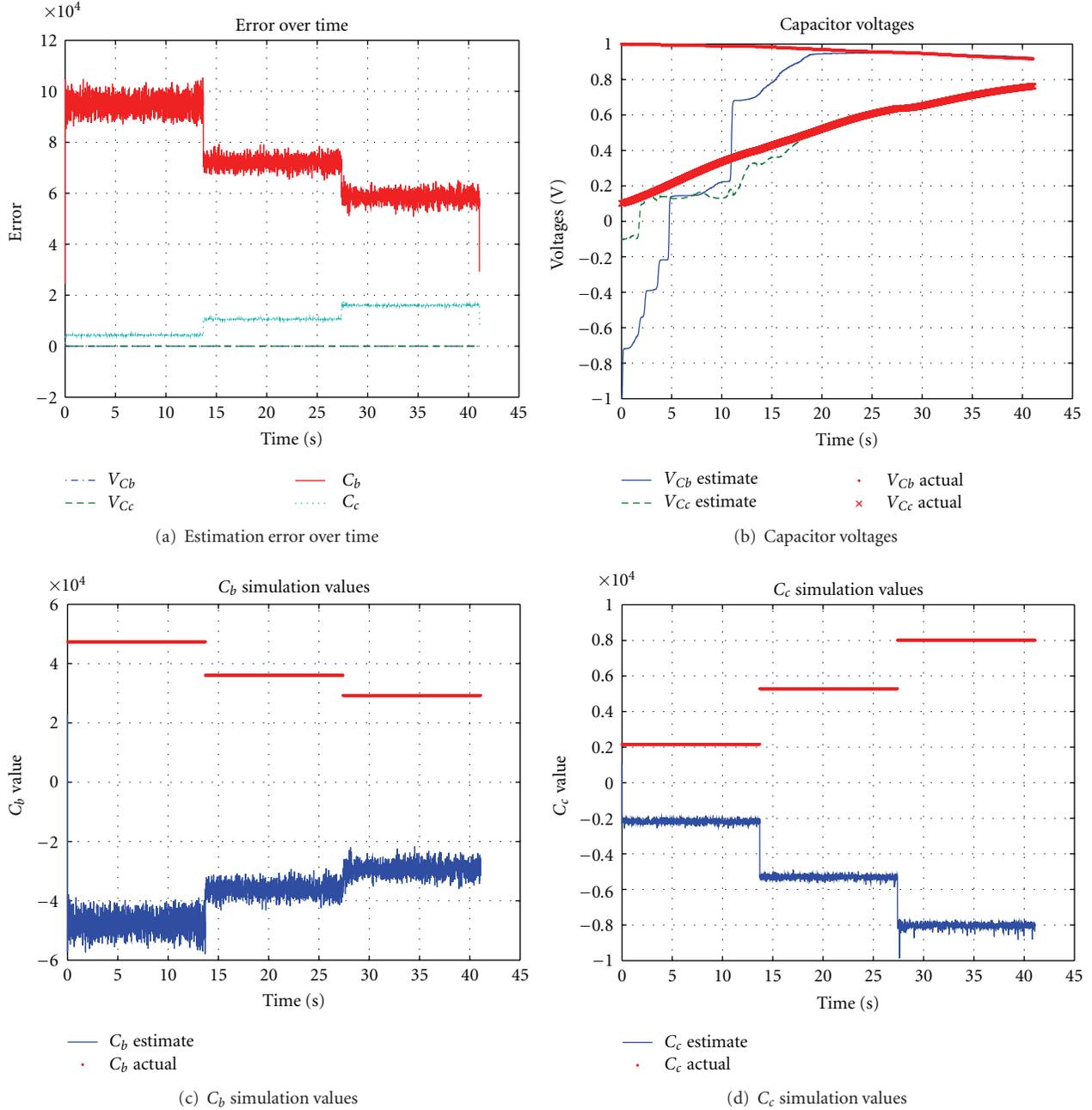

Figure 12: Particle filter simulation results (including errors).

*Time Update Step.*

(1) Assuming at time $k$ the posterior density function $p(x_{k-1} \mid z_{k-1}) = N(\hat{x}_{k-1|k-1}, P_{k-1|k-1})$ is known, then we may factorize as follows:

$$P_{k-1|k-1} = \sqrt{P_{k-1|k-1}} \left(\sqrt{P_{k-1|k-1}}\right)^T. \quad \text{(B.1)}$$

(2) Evaluate the quadrature points $\{X_{l,k-1|k-1}\}_{l=1}^{m}$ as

$$X_{l,k-1|k-1} = \sqrt{P_{k-1|k-1}}\xi_l + \hat{x}_{k-1|k-1}. \quad \text{(B.2)}$$

(3) Evaluate the propagated quadrature points $\{X^*_{l,k|k-1}\}_{l=1}^{m}$ as:

$$X^*_{l,k|k-1} = f(X_{l,k-1|k-1}, u_{k-1}, k-1). \quad \text{(B.3)}$$

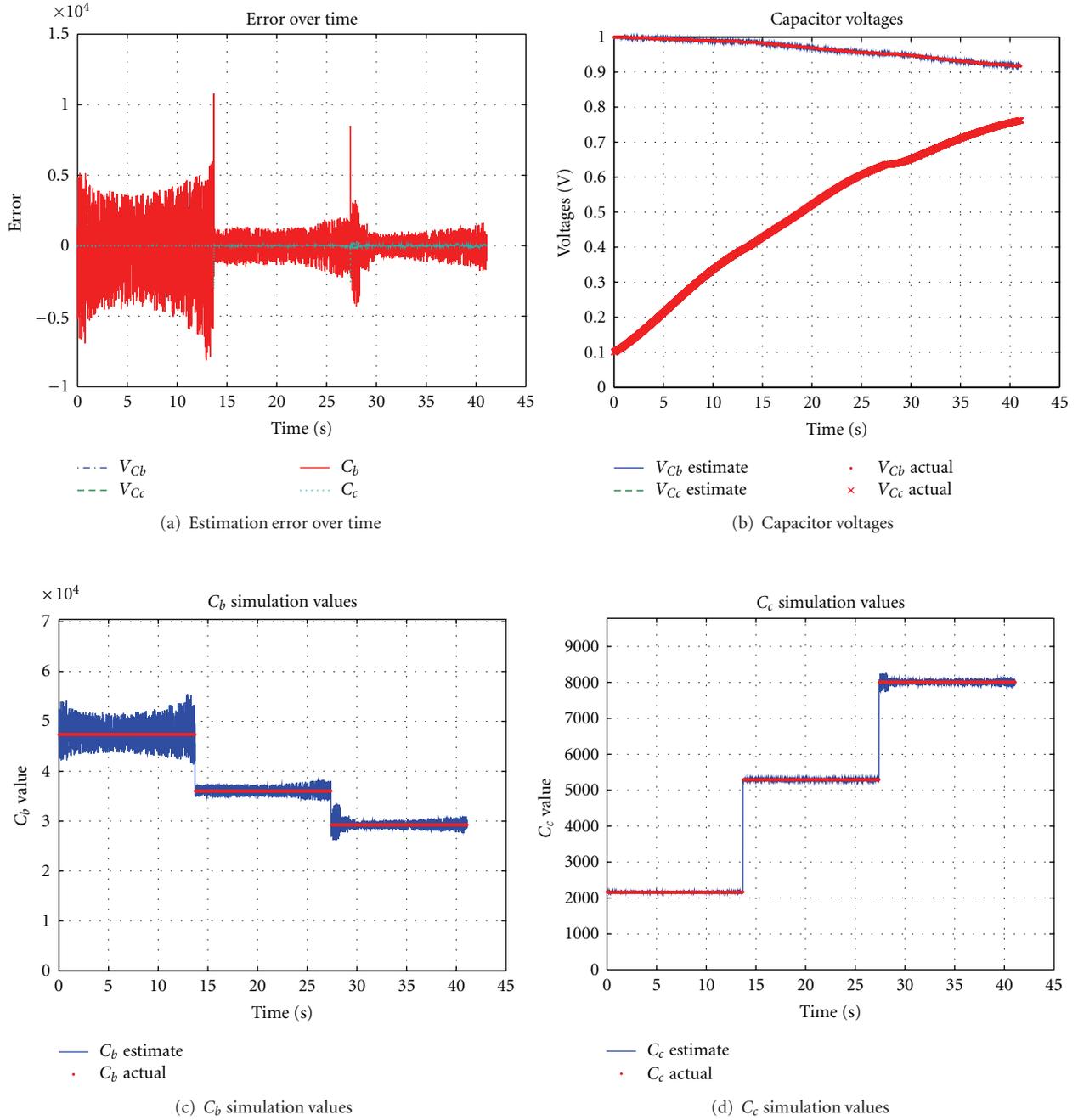

Figure 13: Quadrature Kalman filter simulation results (only noise).

(4) Estimate the predicted state:

$$\hat{x}_{k|k-1} = \sum_{l=1}^{m} w_l X^*_{l,k|k-1}. \tag{B.4}$$

(5) Estimate the predicted error covariance:

$$P_{k|k-1} = \sum_{l=1}^{m} w_l X^*_{l,k|k-1} X^{*T}_{l,k|k-1} - \hat{x}_{k|k-1} \hat{x}^T_{k|k-1} + Q_k. \tag{B.5}$$

*Measurement Update Step.*

(1) Factorize:

$$P_{k-1|k-1} = \sqrt{P_{k-1|k-1}} \left(\sqrt{P_{k-1|k-1}}\right)^T. \tag{B.6}$$

(2) Evaluate the quadrature points $\{X_{l,k-1|k-1}\}_{l=1}^{m}$ as:

$$X_{l,k-1|k-1} = \sqrt{P_{k-1|k-1}} \xi_l + \hat{x}_{k-1|k-1}. \tag{B.7}$$

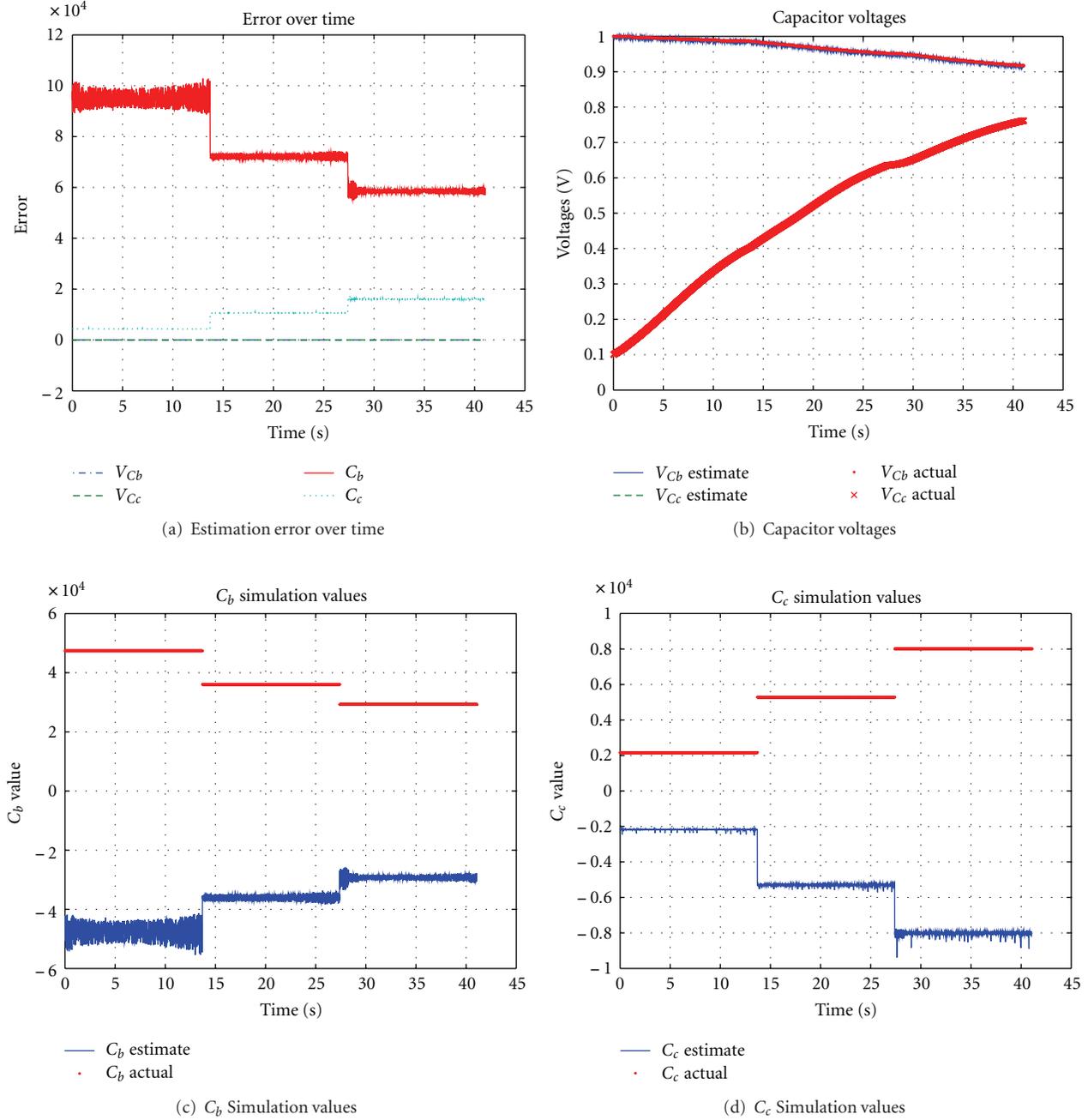

Figure 14: Quadrature Kalman filter simulation results (including errors).

(3) Evaluate the propagated quadrature points $\{Z_{l,k|k-1}\}_{l=1}^{m}$ as:

$$Z_{l,k|k-1} = h(X_{l,k|k-1}, u_k, k). \quad \text{(B.8)}$$

(4) Estimate the predicted measurement:

$$\hat{z}_{k|k-1} = \sum_{l=1}^{m} w_l Z_{l,k|k-1}. \quad \text{(B.9)}$$

(5) Estimate the innovation covariance matrix

$$P_{zz,k|k-1} = R_k + \sum_{l=1}^{m} w_l Z_{l,k|k-1} Z_{l,k|k-1}^T - \hat{z}_{k|k-1} \hat{z}_{k|k-1}^T. \quad \text{(B.10)}$$

(6) Estimate the cross covariance matrix

$$P_{xz,k|k-1} = \sum_{l=1}^{m} w_l X_{l,k|k-1} Z_{l,k|k-1}^T - \hat{x}_{k|k-1} \hat{z}_{k|k-1}^T. \quad \text{(B.11)}$$

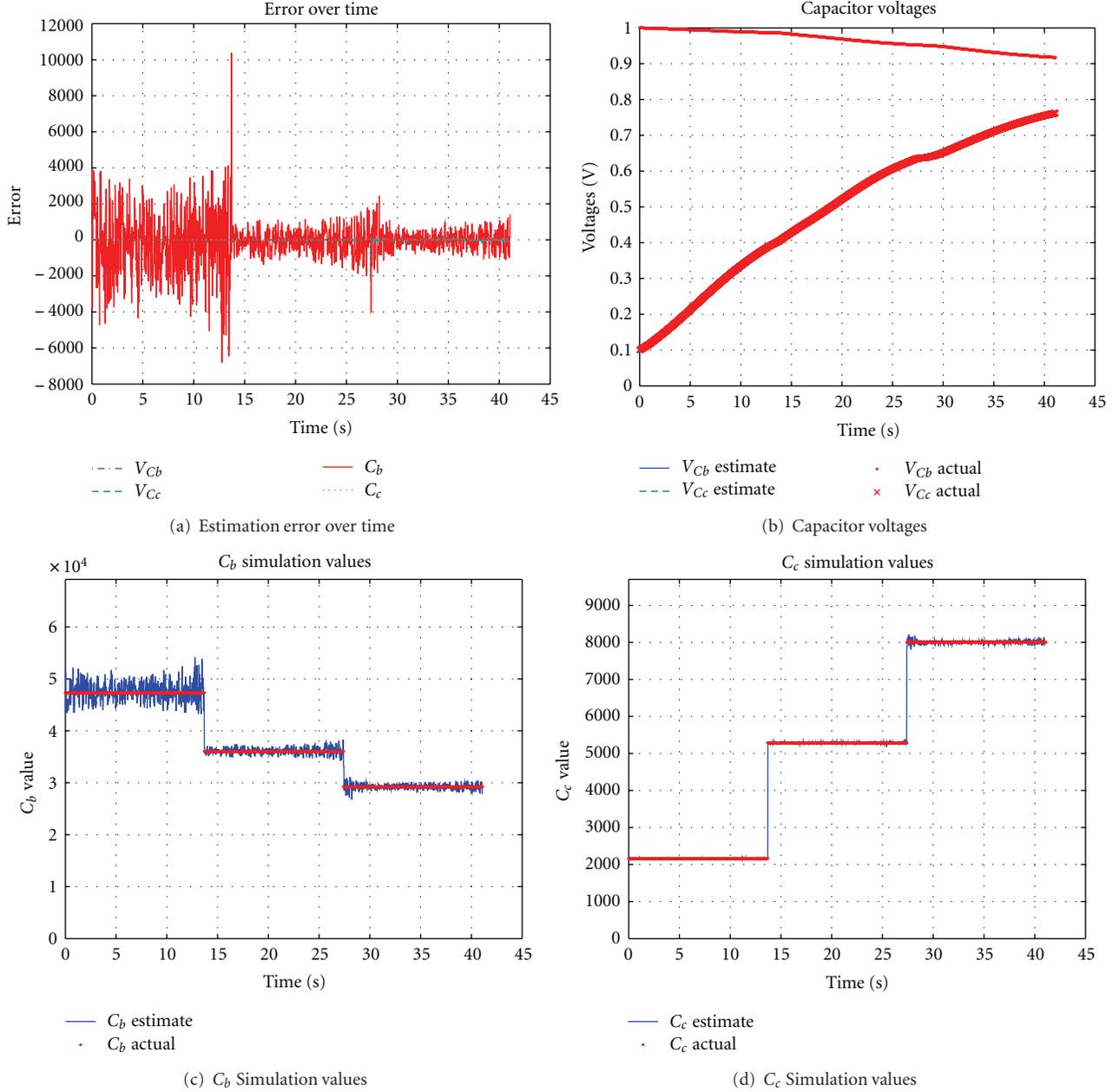

Figure 15: Smooth variable structure filter simulation results (only noise).

(7) Estimate the Kalman gain:

$$W_k = P_{xz,k|k-1} P_{zz,k|k-1}^{-1}. \quad (B.12)$$

(8) Estimate the updated step:

$$\hat{x}_{k|k} = \hat{x}_{k|k-1} + W_k(z_k - \hat{z}_{k|k-1}). \quad (B.13)$$

(9) Estimate the corresponding error covariance:

$$P_{k|k} = P_{k|k-1} - W_k P_{zz,k|k-1} W_k^T. \quad (B.14)$$

## C. Derivation of the RC Battery Model Equations

The output voltage may be calculated by summing the voltages of each element in the circuit. Summation of the outer loop and inner loop voltages yield two equations for the output voltage, respectively:

$$V_O = I_S R_t + I_b R_e + V_{Cb}, \quad (C.1)$$

$$V_O = I_S R_t + I_c R_c + V_{Cc}. \quad (C.2)$$

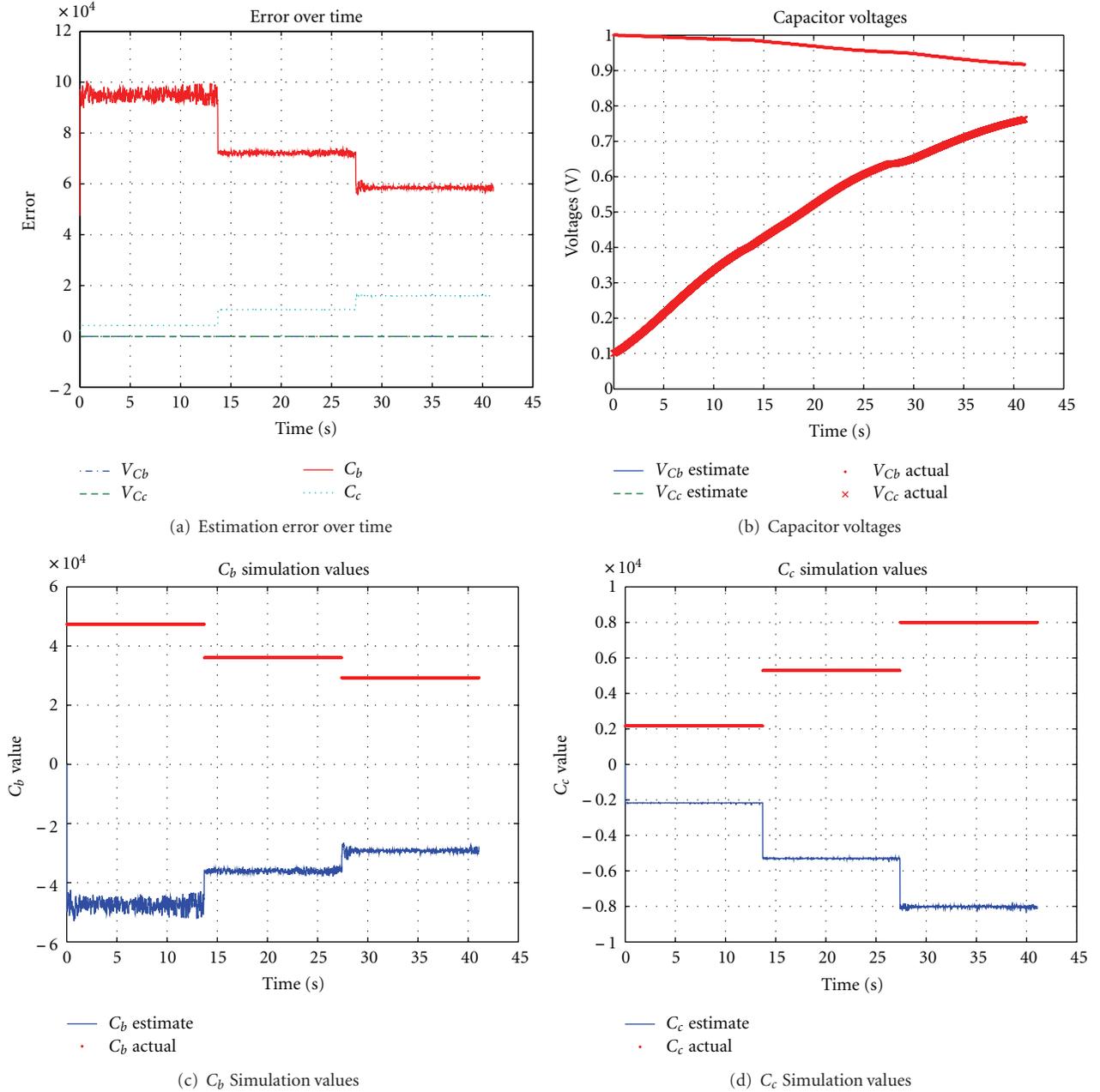

Figure 16: Smooth variable structure filter simulation results (including errors).

Recall Kirchhoff's current law, which states that the total current entering a node must equal the total current leaving, such that the supply current is defined by

$$I_S = I_b + I_c. \quad (C.3)$$

Rearranging for $I_c$ in (C.3) and substituting into (C.2) yields

$$V_O = I_S R_t + (I_S - I_b)R_c + V_{Cc}. \quad (C.4)$$

Equating (C.4) with the first output voltage (C.1) and solving for $I_b$ yields

$$I_b = \frac{I_S R_c}{R_e + R_c} + \frac{V_{Cc}}{R_e + R_c} - \frac{V_{Cb}}{R_e + R_c}. \quad (C.5)$$

Note that $\dot{V}_{Cb} = I_b/C_b$ such that (C.5) becomes

$$\dot{V}_{Cb} = \frac{I_S R_c}{C_b(R_e + R_c)} + \frac{V_{Cc}}{C_b(R_e + R_c)} - \frac{V_{Cb}}{C_b(R_e + R_c)}. \quad (C.6)$$

Similarly, the above approach may be used to solve for the second capacitor voltage rate of change:

$$\dot{V}_{Cc} = \frac{I_S R_e}{C_c(R_e + R_c)} + \frac{V_{Cb}}{C_c(R_e + R_c)} - \frac{V_{Cc}}{C_c(R_e + R_c)}. \quad (C.7)$$

The output voltage may be solved by manipulating (C.1) through (C.3). Setting $I_c = I_S - I_b$ in (C.2), and rearranging for $I_b$, results in the following:

$$I_b = \frac{R_t + R_c}{R_c} I_S + \frac{V_{Cc}}{R_c} - \frac{V_O}{R_c}. \tag{C.8}$$

Rearranging (C.1) for $I_b$, and setting the result equal to (C.8) yields

$$\frac{V_O}{R_e} - \frac{V_{Cb}}{R_e} - \frac{R_t}{R_e} I_S = \frac{R_t + R_c}{R_c} I_S + \frac{V_{Cc}}{R_c} - \frac{V_O}{R_c}. \tag{C.9}$$

Manipulating (C.9) and simplifying for the output voltage results in

$$V_O = \frac{R_c}{R_e + R_c} V_{Cb} + \frac{R_e}{R_e + R_c} V_{Cc} + \left[ R_t + \frac{R_e R_c}{R_e + R_c} \right] I_S. \tag{C.10}$$

## D. Linearization of the Battery System

The battery model described in Section 3 requires linearization for the EKF to be implemented. The linearization was based on the Jacobian matrix and the four states (where for neatness, $W_{Cb} = 1/C_b$ and $W_{Cc} = 1/C_c$)

$$\phi_k = \begin{bmatrix} \frac{\partial f_1}{\partial V_{Cb}} & \frac{\partial f_1}{\partial V_{Cc}} & \frac{\partial f_1}{\partial W_{Cb}} & \frac{\partial f_1}{\partial W_{Cc}} \\ \frac{\partial f_2}{\partial V_{Cb}} & \frac{\partial f_2}{\partial V_{Cc}} & \frac{\partial f_2}{\partial W_{Cb}} & \frac{\partial f_2}{\partial W_{Cc}} \\ \frac{\partial f_3}{\partial V_{Cb}} & \frac{\partial f_3}{\partial V_{Cc}} & \frac{\partial f_3}{\partial W_{Cb}} & \frac{\partial f_3}{\partial W_{Cc}} \\ \frac{\partial f_4}{\partial V_{Cb}} & \frac{\partial f_4}{\partial V_{Cc}} & \frac{\partial f_4}{\partial W_{Cb}} & \frac{\partial f_4}{\partial W_{Cc}} \end{bmatrix}_k, \tag{D.1}$$

where the functions in (D.1) are described by

$$\begin{aligned} f_1 &= -\frac{T_S V_{Cb} W_{Cb}}{R_e + R_c} + V_{Cb} \\ &\quad + \frac{T_S V_{Cc} W_{Cb}}{R_e + R_c} + \frac{T_S R_c W_{Cb}}{R_e + R_c} I_S, \\ f_2 &= -\frac{T_S V_{Cb} W_{Cc}}{R_e + R_c} - \frac{T_S V_{Cc} W_{Cb}}{R_e + R_c} \\ &\quad + V_{Cc} + \frac{T_S R_e W_{Cc}}{R_e + R_c} I_S, \\ f_3 &= W_{Cb}, \\ f_4 &= W_{Cc}. \end{aligned} \tag{D.2}$$

The linearized form ($\phi$) of the system matrix, using the state vector described by (18), is as follows:

$$\phi_k = \begin{bmatrix} \phi_{11} & \phi_{12} & \phi_{13} & 0 \\ \phi_{21} & \phi_{22} & 0 & \phi_{24} \\ 0 & 0 & 1 & 0 \\ 0 & 0 & 0 & 1 \end{bmatrix}_k, \tag{D.3}$$

where the elements are described by:

$$\begin{aligned} \phi_{11} &= \frac{-T_S W_{Cbk}}{R_{ek} + R_{ck}} + 1, \\ \phi_{12} &= \frac{T_S W_{Cbk}}{R_{ek} + R_{ck}}, \\ \phi_{13} &= \frac{T_S}{R_{ek} + R_{ck}} (-V_{Cbk} + V_{Cck} + R_{ck} I_{Sk}), \\ \phi_{21} &= \frac{T_S W_{Cck}}{R_{ek} + R_{ck}}, \\ \phi_{22} &= \frac{-T_S W_{Cck}}{R_{ek} + R_{ck}} + 1, \\ \phi_{24} &= \frac{T_S}{R_{ek} + R_{ck}} (V_{Cbk} - V_{Cck} + R_{ek} I_{Sk}). \end{aligned} \tag{D.4}$$

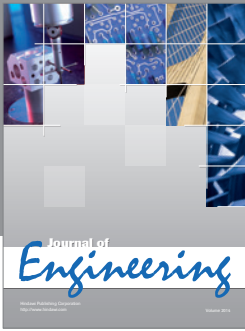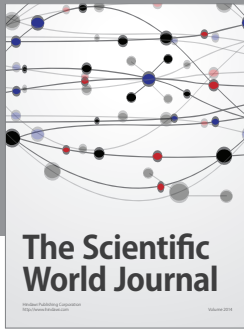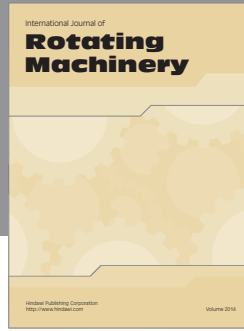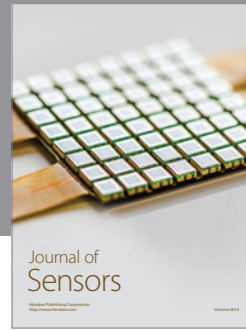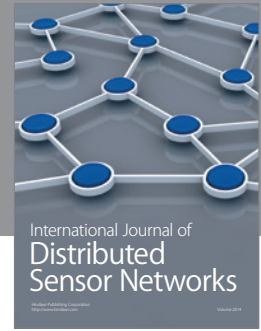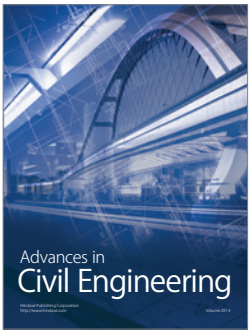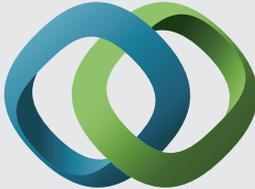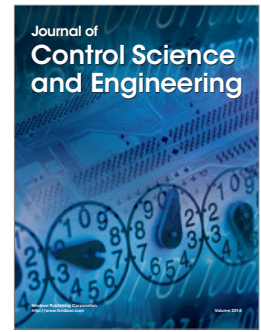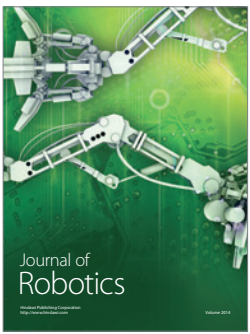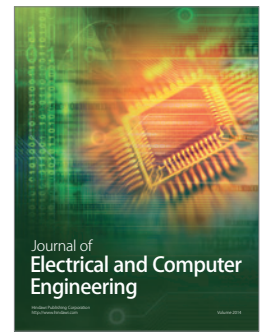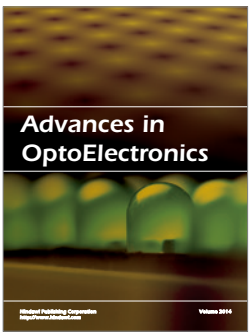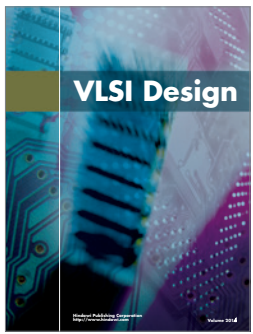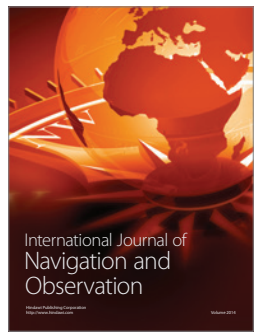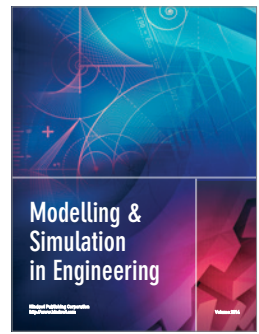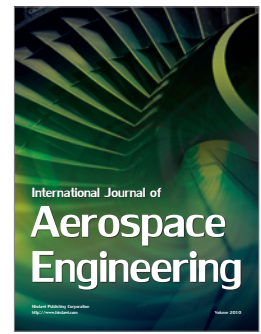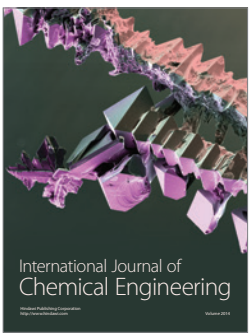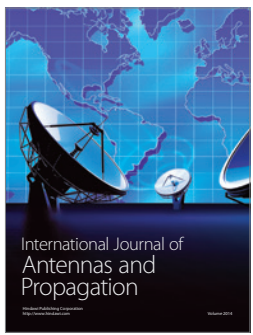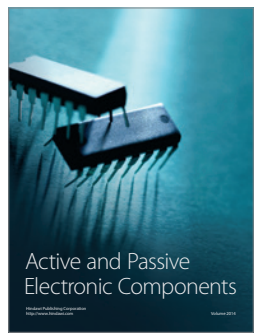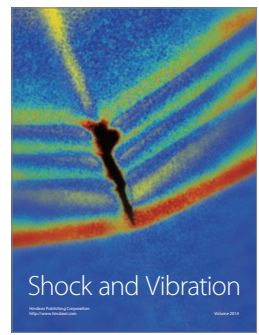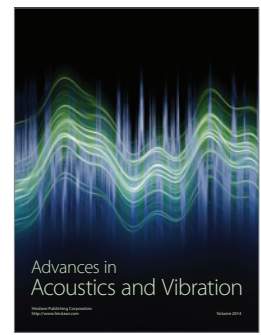